\documentclass[preprint,showpacs,aps,floats,nofootinbib]{revtex4}
\usepackage{amssymb}
\usepackage{graphicx}
\usepackage{dcolumn}
\usepackage{bm}
\usepackage{ulem}
\begin{document}
\newcommand{\bfm}{\bf\boldmath}
\newcommand{\ul}[1]{\underline{{#1}}}
%
%
\title{\boldmath\bf
$K_1(1270)$--$K_1(1400)$ Mixing Angle and
New-Physics Effects in $B \to K_1 \ell^+ \ell^-$ Decays}
\author{Hisaki Hatanaka}
\author{Kwei-Chou Yang}
\affiliation{Department of Physics, Chung-Yuan Christian University,
Chungli 320, Taiwan}
\begin{abstract}
We study semileptonic $B$ meson decays $B \to K_1 (1270) \ell^+ \ell^-$
and $K_1 (1400) \ell^+ \ell^-$ ($\ell \equiv e$, $\mu$, $\tau$), where
the strange $P$-wave mesons, $K_1(1270)$ and $K_1(1400)$, are the
mixtures of the $K_{1A}$ and $K_{1B}$, which are the $1^3P_1$ and
$1^1P_1$ states, respectively.  We show that the ratio $R_\ell \equiv
{\cal B}(B \to K_1(1400)\ell^+\ell^-)/{\cal B}(B \to
K_1(1270)\ell^+\ell^-)$, insensitive to new-physics parameters, is
suitable for determining the $K_1(1270)$--$K_1(1400)$ mixing angle,
$\theta_{K_1}$. The forward-backward asymmetry shows a weak
$\theta_{K_1}$-dependence for $B \to K_1(1270)\mu^+\mu^-$, but
relatively strong for $B\to K_1(1400)\mu^+\mu^-$.  We investigate
model-independent new-physics corrections to operators relevant to the
$b \to s \ell^+ \ell^-$ electroweak-penguin and weak-box
diagrams. Furthermore, for the $B\to K_1(1270)\mu^+\mu^-$ decay the
position of the forward-backward asymmetry zero, which is almost
independent of the value of $\theta_{K_1}$, can be dramatically changed
under variation of new-physics parameters.
\end{abstract}
%
%
\maketitle
\newcommand{\bra}[1]{\langle {#1}}
\newcommand{\ket}[1]{{#1} \rangle}
\newcommand{\ebar}{{\bar{e}}}
\newcommand{\sbar}{\bar{s}}
\newcommand{\cbar}{\bar{c}}
\newcommand{\bbar}{\bar{b}}
\newcommand{\qbar}{\bar{q}}
\renewcommand{\l}{\ell}
\newcommand{\lbar}{\bar{\ell}}
\newcommand{\psibar}{\bar{\psi}}
\newcommand{\barB}{\overline{B}}
\newcommand{\barK}{\overline{K}}
\newcommand{\thetaK}{\theta_{K_1}}
\newcommand{\onepone}{{1^1P_1}}
\newcommand{\sanpone}{{1^3P_1}}
\newcommand{\kone}{{K_1}}
\newcommand{\barkone}{{\overline{K}_1}}
\renewcommand{\Re}{\mathop{\mbox{Re}}}
\renewcommand{\Im}{\mathop{\mbox{Im}}}
\newcommand{\T}{{\cal T}}
\newcommand{\eff}{{\rm eff}}
\newcommand{\A}{{\cal A}}
\newcommand{\B}{{\cal B}}
\newcommand{\C}{{\cal C}}
\newcommand{\D}{{\cal D}}
\newcommand{\E}{{\cal E}}
\newcommand{\F}{{\cal F}}
\newcommand{\G}{{\cal G}}
\renewcommand{\H}{{\cal H}}
\newcommand{\hats}{\hat{s}}
\newcommand{\hatp}{\hat{p}}
\newcommand{\hatq}{\hat{q}}
\newcommand{\hatm}{\hat{m}}
\newcommand{\hatu}{\hat{u}}
\newcommand{\alphaem}{\alpha_{\rm em}}
\newcommand{\konel}{K_1(1270)}
\newcommand{\koneh}{K_1(1400)}
\newcommand{\barkonel}{\barK_1(1270)}
\newcommand{\barkoneh}{\barK_1(1400)}
\newcommand{\konea}{K_{1A}}
\newcommand{\koneb}{K_{1B}}
\newcommand{\barkonea}{\barK_{1A}}
\newcommand{\barkoneb}{\barK_{1B}}
\newcommand{\mkone}{m_{\kone}}
\newcommand{\konep}{K_1^+}
\newcommand{\konem}{K_1^-}
\newcommand{\konelm}{K_1^-(1270)}
\newcommand{\konehm}{K_1^-(1400)}
\newcommand{\konelp}{K_1^+(1270)}
\newcommand{\konehp}{K_1^+(1400)}
\newcommand{\konelz}{\overline{K}{}^0_1(1270)}
\newcommand{\konehz}{\overline{K}{}^0_1(1400)}
\newcommand{\Bm}{B^-}
\newcommand{\Bz}{\overline{B}{}^0}
\newcommand{\Kstar}{K^*(892)}
\newcommand{\BABAR}{BABAR}
\newcommand{\BELLE}{Belle}
\newcommand{\CLEO}{CLEO}
\newcommand{\leftu}{\gamma^\mu L}
\newcommand{\leftd}{\gamma_\mu L}
\newcommand{\rightu}{\gamma^\mu R}
\newcommand{\rightd}{\gamma_\mu R}
\newcommand{\Br}{{\cal B}}
\newcommand{\sect}[1]{Sec.~\ref{#1}}
\newcommand{\eqref}[1]{(\ref{#1})}
\newcommand{\fig}{FIG.~}
\newcommand{\figs}{FIGs.~}
\newcommand{\tbl}{TABLE~}
\newcommand{\tbls}{TABLEs~}
\newcommand{\errpm}[3]{#1^{+{#2}}_{-{#3}}}
\newcommand{\errpmf}[5]{{#1}^{ +{#2} +{#4} }_{-{#3}-{#5}}}
\newcommand{\lpm}{\l^+\l^-}
\newcommand{\epm}{e^+e^-}
\newcommand{\mupm}{\mu^+\mu^-}
\newcommand{\taupm}{\tau^+\tau^-}
\newcommand{\AFB}{A_{\rm FB}}
\newcommand{\barAFB}{\overline{A}_{\rm FB}}
\newcommand{\GeV}{{\,\mbox{GeV}}}
\newcommand{\MeV}{{\,\mbox{MeV}}}
\newcommand{\degree}{^\circ}
\newcommand{\x}{\phantom{0}}
\newcommand{\mB}{m_B}
\newcommand{\SM}{{\rm SM}}
\newcommand{\NP}{{\rm NP}}
\newcommand{\barc}{\bar{c}}
\newcommand{\xipara}{\xi_\parallel^{\kone}}
\newcommand{\xiperp}{\xi_\perp^{\kone}}
\newcommand{\xiparal}{\xi_\parallel^{\konel}}
\newcommand{\xiperpl}{\xi_\perp^{\konel}}
\newcommand{\xiparah}{\xi_\parallel^{\koneh}}
\newcommand{\xiperph}{\xi_\perp^{\koneh}}
\newcommand{\para}{\parallel}
\newcommand{\alphas}{\alpha_s}
\newcommand{\pA}{p_{\kone}}
\newcommand{\lcaption}[2]{\caption{(label:{#2}) #1}\label{#2}}
\newcommand{\Rmunr}{R_{\mu,\rm nr}}
\newcommand{\RdGamma}{R_{d\Gamma/ds,\mu}}
\providecommand{\dfrac}[2]{\frac{\displaystyle {#1}}{\displaystyle{#2}}}
%
\section{Introduction}
%
\begin{table}[tbp]
\caption{ Experimental status of branching fractions (in units of
$10^{-6}$) for the decays $B \to \Kstar \gamma$, $\konel\gamma,
\koneh\gamma$ and $B \to \Kstar \l^+\l^-$ \cite{HFAG}.
}\label{kll-data}
\begin{center}
\begin{ruledtabular}
\begin{tabular}{lcl|lcl}
Mode & Exp.(Average) & Ref. &
Mode & Exp.(Average) & Ref. \\
\hline
$K^{*+}(892) \gamma$ & $40.3 \pm 2.6$ &
 \cite{Aubert:2004te,Nakao:2004th,Coan:1999kh} &
$K^{*0}(892) \gamma$ & $40.1 \pm 2.0$ &
 \cite{Aubert:2004te,Nakao:2004th,Coan:1999kh}
\\
$K_1^+(1270) \gamma$ & $43\pm12$ & \cite{Yang:2004as} &
$K_1^0(1270) \gamma$ & $<58$     & \cite{Yang:2004as}
\\
$K_1^+(1400) \gamma$ & $<15$     & \cite{Yang:2004as} &
$K_1^0(1400) \gamma$ & $<15$     & \cite{Yang:2004as}
\\
$K^{*+}(892) \epm$ & $1.23^{+0.69}_{-0.62}$ &
 \cite{Ishikawa:2003cp,Aubert:2006vb} &
$K^{*0}(892) \epm$ & $1.11^{+0.30}_{-0.26}$ &
 \cite{Ishikawa:2003cp,Aubert:2006vb}
\\
$K^{*+}(892) \mupm$ & $0.78^{+0.56}_{-0.44}$ &
 \cite{Ishikawa:2003cp,Aubert:2006vb} &
$K^{*0}(892) \mupm$ & $0.98^{+0.22}_{-0.21}$ &
 \cite{Ishikawa:2003cp,Aubert:2006vb}
\\
\end{tabular}
\end{ruledtabular}
\end{center}
\end{table}
%
%
$b \to s$ transitions in semileptonic and radiative $B$ meson decays
contain rich phenomena relevant to the standard model (SM) and new
physics (NP).  Semileptonic and radiative $B$ decays involving a vector
or axial vector meson have been observed by \BABAR, \BELLE\ and \CLEO\
(see Table~\ref{kll-data}).  The rare flavor-changing neutral-current
processes, $b\to s \bar{\ell} \ell$, which proceed through the
electroweak-penguin and weak-box diagrams in the SM, may provide a
hunting ground to search for the NP effects. For $B\to \Kstar\lpm$
decays, the forward-backward asymmetry has been measured by
\BABAR~\cite{Aubert:2006vb} and \BELLE~\cite{Ishikawa:2006fh}. Very
recently, BABAR \cite{aubert:2008ju,Aubert:2008ps,Eigen:2008nz} has
reported the measurements for the longitudinal polarization fraction and
forward-backward asymmetry (FBA) of $B\to K^*(892) \ell^+ \ell^-$, and
for the isospin asymmetry of $B^0\to K^{*0}(892) \ell^+ \ell^-$ and
$B^\pm\to K^{*\pm}(892) \ell^+ \ell^-$ channels. The data may hint at
the flipped sign(s) of the Wilson coefficients, {\it e.g.}, the flipped
sign of $c^{\rm eff}_7$ related to the magnetic dipole operator. To
extract the moduli and arguments of the effective Wilson coefficients,
it is important to measure various observables in different inclusive
and exclusive rare processes. These should be considerably improved at
LHCb.

The radiative $B$ decay involving the $\konel$, the orbitally excited
($P$-wave) state, is recently observed by \BELLE\ and other radiative and
semileptonic decay modes involving $\konel$ and $\koneh$ are hopefully expected
to be seen soon. Some studies for $B\to\kone\lpm$ have been made recently
\cite{Paracha:2007yx,Ahmed:2008ti,Saddique:2008xj}.  Just like $B \to
\Kstar\lpm$ decays
\cite{Ali:1999mm,Ali:2002jg,Beneke:2001at,Feldmann:2002iw,Kruger:2005ep,Bobeth:2008ij,Egede:2008uy,Chen:2008ug},
$B\to\kone\lpm$ decays can offer the good probe to the NP, and are much more
sophisticated due to the mixing of the $\konea$ and $\koneb$, which are the
$\sanpone$ and $\onepone$ states, respectively.  The physical $K_1$ mesons are
$\konel$ and $\koneh$, described by
\begin{eqnarray}
\pmatrix{|\barkonel \rangle \cr |\barkoneh \rangle}
= M \pmatrix{| \barkonea \rangle \cr | \barkoneb \rangle},
\quad\mbox{with} \quad
M = \pmatrix{
\sin \thetaK & \phantom{-} \cos \thetaK \cr
\cos \thetaK & -\sin \thetaK}. \label{mixing}
\end{eqnarray}
The magnitude of $\thetaK$ was estimated to be $|\thetaK| \approx
34\degree\vee 57\degree$ in Ref. \cite{Suzuki:1993yc}, $35\degree \lesssim
|\thetaK| \lesssim 55\degree$ in Ref.~\cite{Burakovsky:1997ci}, and
$|\thetaK|= 37\degree \vee 58\degree$ in Ref.~\cite{Cheng:2003bn}.
Nevertheless, the sign of the $\thetaK$ was not yet determined in these
studies.  From the study for $B\to\konel\gamma$ and
$\tau\to\konel\nu_\tau$, we recently obtain \cite{Hatanaka:2008xj}
\begin{eqnarray}
\thetaK= -(34 \pm 13)\degree, \label{thetaKvalue}
\end{eqnarray}
where the minus sign of $\thetaK$ is related to the chosen phase of
$|\ket{\barkonea}$ and $|\ket{\barkoneb}$. We adopt the following conventions
\cite{Hatanaka:2008xj}: $f_{\konea} > 0$ and $f_{\koneb}^\perp > 0$, which are
defined by
\begin{eqnarray}
\bra{0}|\psibar \gamma^\mu \gamma_5 s |\ket{\barkonea (P,\lambda)}
&=&
- i f_{\konea} m_{\konea} \varepsilon_\mu^{(\lambda)},
\nonumber \\
\bra{0}|\psibar \sigma_{\mu\nu} s |\ket{\barkoneb (P,\lambda)}
&=&
  i f^{\perp}_{\koneb} \epsilon_{\mu\nu\alpha\beta}
\varepsilon_{(\lambda)}^\alpha
P^\beta,\quad
\psi \equiv d,u.
\end{eqnarray}
Within the SM, we have predicted \cite{Hatanaka:2008xj}
\begin{eqnarray}
{\cal B}(B^-\to K_1^-(1270)\gamma)&=&(66^{+50}_{-30}) \times 10^{-6}
\left(\frac{m_{b.pole}}{4.90~{\rm GeV}}\right)^2~,
\\ {\cal B}(B^-\to K_1^-(1400)\gamma)&=&(6.5^{+12.8}_{-~6.3})
\times 10^{-6} \left(\frac{m_{b.pole}}{4.90~{\rm GeV}}\right)^2~,
\end{eqnarray}
where $m_{b,pole}$ is the pole mass of the $b$ quark. In the present paper, we
study the observables for $B \to \kone \lpm$ decays, including the dilepton
mass spectra, decay rates and forward-backward asymmetries. We further show
that the mixing angle $\thetaK$ can be determined from the $B\to\kone\lpm$
decays. In addition to the study of the $\thetaK$, we also investigate the
model-independent new-physics corrections to the Wilson coefficients
$c_7^\eff$, $c_9$ and $c_{10}$.  The new-physics parameters can be well
constrained by the measurement of $B\to\kone\lpm$ forward-backward asymmetry
(FBA), where the position of the FBA zero depends very weakly on the value of
the $\thetaK$. Hence, the position of zero of the differential FBAs depends on
the underlying new physics corrections.

This paper is organized as follows. In \sect{sec:Hamiltonian} we introduce the
effective Hamiltonian and effective operators therein. In \sect{sec:mixing}, we
give the definitions for $B\to\konel$ and $B\to\koneh$ form factors. In
\sect{sec:decay}, we formulate the $B \to \kone\lpm$ decays and discuss
determination of the $\thetaK$ in details. In \sect{sec:NP}, we estimate the NP
effects in the model-independent way. We summarize the main results in
\sect{sec:summary}.

%
\section{The effective Hamiltonian}\label{sec:Hamiltonian}
%
Neglecting doubly Cabibbo-suppressed contributions, the effective weak
Hamiltonian relevant to $b\to s \lpm$ is given by
\begin{equation}
{\cal H}_{\rm eff}= -4\frac{G_F}{\sqrt{2}}V_{tb}^{ }V_{ts}^* \sum_{i=1}^{10}
c_i(\mu)O_i(\mu)~,
\end{equation}
where the Wilson operators $O_i$ for $i=1,\cdots,10$ read
\cite{Buras:1993xp}
\begin{eqnarray}
O_1 &=&
 (\sbar_\alpha \leftd c_\alpha)(\cbar_\beta \leftu b_\beta),
\quad
O_2  =
 (\sbar_\alpha \leftd c_\beta)(\cbar_\beta \leftu b_\alpha),
\nonumber\\
O_3 &=& \textstyle (\sbar_\alpha \leftd b_\alpha) \sum_q
      (\qbar_\beta \leftu q_\beta),
\quad
O_4  = \textstyle (\sbar_\alpha \leftd b_\beta) \sum_q
      (\qbar_\beta \leftu q_\alpha),
\nonumber\\
O_5 &=& \textstyle (\sbar_\alpha \leftd  b_\alpha) \sum_q
      (\qbar_\beta \rightu q_\beta),
\quad
O_6 = \textstyle (\sbar_\alpha \leftd  b_\beta) \sum_q
      (\qbar_\beta \rightu q_\alpha),
\nonumber\\
O_7&=&
 \frac{e m_b}{16\pi^2}
 \sbar \sigma^{\mu\nu} R b F_{\mu\nu},
\nonumber\\
O_9 &=&
\frac{\alphaem}{4\pi}(\lbar \gamma_\mu \l) (\sbar\gamma^\mu L b),
\quad
O_{10}  =
\frac{\alphaem}{4\pi}(\lbar \gamma_\mu\gamma_5 \l)
  (\sbar\gamma^\mu L b),
\end{eqnarray}
with $L = (1-\gamma_5)/2$, $R = (1+\gamma_5)/2$, and $\alpha,\beta$
being the $SU(3)$ color indices.
%
\begin{table}
\caption{ The Wilson coefficients $c_i(\mu)$ at the scale $\mu =
m_{b,{\rm pole}}$   in the SM. Here $c_7^\eff \equiv c_7 -
\frac{1}{3}c_5 - c_6$.  }\label{WC}
\begin{center}
\begin{ruledtabular}
\begin{tabular}{cccccccccc}
$\barc_1$    & $\barc_2$    & $\barc_3$    & $\barc_4$    & $\barc_5$ &
$\barc_6$    & $c_7^\eff$   & $c_9$        & $c_{10}$
\\
\hline
$+1.107$ & $-0.248$ & $-0.011$ & $-0.026$ & $-0.007$ &
$-0.031$ & $-0.313$ & $ 4.344$ & $-4.669$
\end{tabular}
\end{ruledtabular}
\end{center}
\end{table}
%
The $b \to s \l^+\l^-$ decay amplitude is given by
\begin{eqnarray}
{\cal M}(b \to s \l^+  \l^-)
&=&
\frac{G_F}{\sqrt{2}}
\frac{\alphaem}{\pi}
V_{ts}^* V_{tb}^{}
\Bigg\{
  c_9^\eff(\hats)   [\sbar \gamma_\mu L b] [\lbar \gamma^\mu \l]
+ c_{10}[\sbar \gamma_\mu L b] [\lbar \gamma^\mu \l]
\nonumber\\
&& -2\hat{m}_b c_{7}^\eff
\left[
 \sbar i \sigma_{\mu\nu} \frac{\hat{q}^\nu}{\hat{s}} R b
\right]
[\lbar \gamma^\mu \l]
\Bigg\},
\label{eq:amplitude}
\end{eqnarray}
where $\hatm_b \equiv \bar{m}_b / m_B$ with $\bar m_b=\bar m_b(\bar m_b)$ being the $b$ quark mass in the $\overline{\rm MS}$ scheme, $\hats = q^2/m_B^2$, $q_\mu = (p_+ + p_-)_\mu$ with
$p_\pm$ being momenta of the leptons $\l^{\pm}$. To next-to-leading order the running $\overline{\rm MS}$ and pole $b$-quark masses are related by
\begin{equation}
 \overline{m}_b (\mu) = m_{b,\rm pole} \left[ 1 - \frac{\alpha_s(\mu) C_F}{4\pi}
  \left( 4 - 3\ln\frac{m_{b,\rm pole}^2}{\mu^2}\right)
  + {\cal O}(\alpha_s^2 )\right]\,,
\end{equation}
where $C_F=(N_c^2 -1)/(2N_c)$ with $N_c$ being the number of colors. In Eq.~\eqref{eq:amplitude}
we have neglected ${\cal O}(m_s/m_b)$ corrections. $c_9^\eff(\hats) = c_9 +
Y(\hats)$, where $Y(\hats) = Y_{\rm pert}(\hats) + Y_{\rm LD}$ contains both
the perturbative part $Y_{\rm pert}(\hats)$ and long-distance part $Y_{\rm
LD}(\hats)$.
$Y(\hats)_{\rm pert}$ is given by \cite{Buras:1994dj}
\begin{eqnarray}
Y_{\rm pert} (\hats) &=&
g(\hatm_c,\hats) c_0
\nonumber\\&&
-\frac{1}{2} g(1,\hats) (4 \barc_3 + 4 \barc_4 + 3 \barc_5 + \barc_6)
-\frac{1}{2} g(0,\hats) (\barc_3 + 3 \barc_4)
\nonumber\\&&
+\frac{2}{9} (3 \barc_3 + \barc_4 + 3 \barc_5 + \barc_6),
\\
\mbox{with}\quad c_0
&\equiv& \barc_1 + 3\barc_2 + 3 \barc_3 + \barc_4 + 3 \barc_5 + \barc_6,
\end{eqnarray}
and the function $g(x,y)$ defined in \cite{Buras:1994dj}.  Here $\barc_1$ --
$\barc_6$ are the Wilson coefficients in the leading logarithmic approximation.
The relevant Wilson coefficients are collected in Table~\ref{WC}
\cite{Buras:1993xp,Ali:1999mm}. $Y(\hats)_{\rm LD}$ involves $B \to K_1 V(\cbar
c)$ resonances \cite{Ali:1991is,Lim:1988yu,Kruger:1996cv}, where $V(\cbar c)$
are the vector charmonium states. We follow Refs.~\cite{Ali:1991is,Lim:1988yu}
and set
\begin{eqnarray}
Y_{\rm LD}(\hats)
&=&
 - \frac{3\pi}{\alphaem^2} c_0
\sum_{V = \psi(1s),\cdots}
\kappa_V \frac{\hatm_V \Br(V\to \l^+\l^-)\hat{\Gamma}_{\rm tot}^V}{\hats - \hatm_V^2 + i \hatm_V \hat{\Gamma}_{\rm tot}^V},
\end{eqnarray}
where $\hat{\Gamma}_{\rm tot}^V \equiv \Gamma_{\rm tot}^V/\mB$ and $\kappa_V
=2.3$. The relevant properties of vector charmonium states are summarized in
Table~\ref{charmonium}.
\begin{table}[tbp]
\caption{Masses, total decay widths and branching fractions of dilepton
decays of vector charmonium states \cite{Yao:2006px}.}\label{charmonium}
\begin{center}
\begin{ruledtabular}
\begin{tabular}{cclll}
$V$ & Mass[\GeV] &  $\Gamma_{\rm tot}^V$[\MeV]
 &\multicolumn{2}{c}{$\Br(V\to\lpm)$}
\\
\hline
$J/\Psi(1S)$ & $3.097$ & $\x\x0.093$ & $5.9\times10^{-2}$ & for $\l=e,\mu$
\\
$\Psi(2S)$   & $3.686$ & $\x\x0.327$ & $7.4\times10^{-3}$ & for $\l=e,\mu$ \\
             &         &             & $3.0\times10^{-3}$ & for $\l=\tau$
\\
$\Psi(3770)$ & $3.772$ & $\x25.2$ & $9.8\times10^{-6}$ & for $\l=e$
\\
$\Psi(4040)$ & $4.040$ & $\x80$ & $1.1\times10^{-5}$ & for $\l=e$
\\
$\Psi(4160)$ & $4.153$ & $103$ & $8.1\times10^{-6}$ & for $\l=e$
\\
$\Psi(4415)$ & $4.421$ & $\x62$ & $9.4\times10^{-6}$ & for $\l=e$
\end{tabular}
\end{ruledtabular}
\end{center}
\end{table}
%
\section{$B\to\konel$ and $B\to\koneh$ form factors}\label{sec:mixing}
%
The $\barB(p_B)\to \barkone(\pA,\lambda)$ form factors are defined by
\begin{eqnarray}
\lefteqn{\bra{\barkone(\pA,\lambda)}|\psibar \gamma_\mu
(1-\gamma_5) b|\ket{\barB(p_B)}}
\quad&&\nonumber\\
&=&
 -i \frac{2}{m_B + \mkone} \epsilon_{\mu\nu\rho\sigma}
\varepsilon_{(\lambda)}^{*\nu} p_B^\rho \pA^\sigma A^{\kone}(q^2)
\nonumber
\\
&& -\left[
(m_B + \mkone)\varepsilon_\mu^{(\lambda)*} V_1^{\kone}(q^2)
-
(p_B + \pA)_\mu (\varepsilon_{(\lambda)}^* \cdot p_B)
 \frac{V_2^{\kone}(q^2)}{m_B + \mkone}
\right]
\nonumber\\&&
+2 \mkone \frac{\varepsilon_{(\lambda)}^* \cdot p_B}{q^2} q_\mu
\left[
 V_3^{\kone}(q^2) - V_0^{\kone}(q^2)
\right],
\label{formfactor1}
\\
\lefteqn{\bra{\barkone(\pA,\lambda)}|\psibar \sigma_{\mu\nu}
q^\nu (1+\gamma_5) b|\ket{\barB(p_B)}}
\quad&&\nonumber\\
&=&
2T_1^{\kone}(q^2) \epsilon_{\mu\nu\rho\sigma}
\varepsilon_{(\lambda)}^{*\nu} p_B^\rho  \pA^\sigma
\nonumber\\
&& -i T_2^{\kone}(q^2)
\left[
 (m_B^2 - \mkone^2) \varepsilon^{(\lambda)}_{*\mu}
-(\varepsilon_{(\lambda)}^{*}\cdot q)
 (p_B + \pA)_\mu
\right]
\nonumber\\&&
- iT_3^{\kone}(q^2) (\varepsilon_{(\lambda)}^{*} \cdot q)
\left[
q_\mu - \frac{q^2}{m_B^2 - \mkone^2} (\pA + p_B)_\mu
\right],
\label{formfactor2}
\end{eqnarray}
where $q \equiv p_B - \pA$, $\gamma_5 \equiv
i\gamma^0\gamma^1\gamma^2\gamma^3$, $\epsilon^{0123} = -1$, and
$\psi\equiv d$, $s$.  The form factors satisfy the following relations,
\begin{eqnarray}
V_3^\kone(0) &=& V_0^{\kone}(0),
\quad
T_1^{\kone}(0) = T_2^{\kone}(0),
\nonumber \\
V_3^\kone(q^2) &=& \frac{m_B + m_{\kone}}{2 m_{\kone}} V_1^\kone(q^2)-
           \frac{m_B - m_{\kone}}{2 m_{\kone}} V_2^\kone(q^2).
\end{eqnarray}
Because the $\konel$ and $\koneh$ are the mixing states of the $\konea$
and $\koneb$, the $\barB \to \barK_1$ form factors can be parametrized
by
\begin{eqnarray}
\pmatrix{
 \bra{\barkonel}|\sbar \gamma_\mu(1-\gamma_5) b |\ket{\barB} \cr
 \bra{\barkoneh}|\sbar \gamma_\mu(1-\gamma_5) b |\ket{\barB}}
&=&
M
\pmatrix{
 \bra{\barK_{1A}}|\sbar\gamma_\mu(1-\gamma_5) b|\ket{\barB} \cr
 \bra{\barK_{1B}}|\sbar\gamma_\mu(1-\gamma_5) b|\ket{\barB} },
\\
\pmatrix{
\bra{\barkonel}|\sbar \sigma_{\mu\nu}q^\nu(1+\gamma_5) b |\ket{\barB} \cr
\bra{\barkoneh}|\sbar \sigma_{\mu\nu}q^\nu(1+\gamma_5) b |\ket{\barB} }
&=&
M \pmatrix{
 \bra{\barK_{1A}}|\sbar\sigma_{\mu\nu}q^\nu(1+\gamma_5) b|\ket{\barB} \cr
 \bra{\barK_{1B}}|\sbar\gamma_{\mu\nu}q^\nu(1+\gamma_5) b|\ket{\barB} }
,
\end{eqnarray}
with the mixing matrix $M$ being given in Eq.~\eqref{mixing}.  Thus the form
factors $A^\kone,V_{0,1,2}^\kone$ and $T_{1,2,3}^\kone$ satisfy following
relations:
\begin{eqnarray}
\pmatrix{
 A^{\konel}/(m_B + m_{\konel}) \cr
 A^{\koneh}/(m_B + m_{\koneh})}
&=& M
\pmatrix{
 A^{\konea}/(m_B + m_{\konea}) \cr
 A^{\koneb}/(m_B + m_{\koneb})},
\\
\pmatrix{
(m_B+m_{\konel}) V_1^{K_1(1270)} \cr
(m_B+m_{\koneh}) V_1^{K_1(1400)}}
&=& M
\pmatrix{
(m_B+m_{\konea})V_1^{K_{1A}} \cr
(m_B+m_{\koneb})V_1^{K_{1B}}},
\\
\pmatrix{
V_2^{K_1(1270)}/(m_B + m_{\konel}) \cr
V_2^{K_1(1400)}/(m_B + m_{\koneh})}
&=& M
\pmatrix{
V_2^{K_{1A}}/(m_B + m_{\konea}) \cr
V_2^{K_{1B}}/(m_B + m_{\koneb})},
\\
\pmatrix{
m_{\konel} V_0^{K_1(1270)} \cr
m_{\koneh} V_0^{K_1(1400)}}
&=& M
\pmatrix{
m_{\konea} V_0^{K_{1A}} \cr
m_{\koneb} V_0^{K_{1B}}},
\\
\pmatrix{
T_1^{K_1(1270)} \cr
T_1^{K_1(1400)}}&=& M
\pmatrix{
T_1^{K_{1A}} \cr
T_1^{K_{1B}} },
\\
\pmatrix{
(m_B^2 - m_{\konel}^2) T_2^{K_1(1270)} \cr
(m_B^2 - m_{\koneh}^2) T_2^{K_1(1400)}}&=& M
\pmatrix{
(m_B^2 - m_{\konea}^2) T_2^{K_{1A}} \cr
(m_B^2 - m_{\koneb}^2) T_2^{K_{1B}} },
\\
\pmatrix{
T_3^{K_1(1270)} \cr
T_3^{K_1(1400)}}
&=& M
\pmatrix{
T_3^{K_{1A}} \cr
T_3^{K_{1B}} },
\end{eqnarray}
where we have assumed that $p^\mu_{\konel,\koneh} \simeq p^\mu_{\konea} \simeq
p^\mu_{\koneb}$. For the numerical analysis, we use the light-cone sum rule
(LCSR) results for the form factors \cite{Yang:2008xw,yang:tensor-form-factors}
which are exhibited in Table~\ref{tab:FFinLF}, where the momentum dependence is
parametrized in the three-parameter form:
\begin{eqnarray} \label{eq:FFpara}
 F(q^2)=\,{F(0)\over 1-a(q^2/m_{B}^2)+b(q^2/m_{B}^2)^2}.
\end{eqnarray}
\begin{table}[t]
\caption{Form factors for $B\to K_{1A},K_{1B}$ transitions obtained in the LCSR
calculation \cite{Yang:2008xw,yang:tensor-form-factors} are fitted to the
3-parameter form in Eq. (\ref{eq:FFpara}).} \label{tab:FFinLF}
\begin{ruledtabular}
\begin{tabular}{clll|clll}
      ~~~~$F$~~~~~~
    & ~~~~~$F(0)$~~~~~
    & ~~~$a$~~~
    & ~~~$b$~~
    & ~~~~$F$~~~~~~
    & ~~~~~$F(0)$~~~~~
    & ~~~$a$~~~
    & ~~~$b$~~
 \\
    \hline
$V_1^{BK_{1A}}$
    & $0.34\pm0.07$
    & $0.635$
    & $0.211$
&$V_1^{BK_{1B}}$
    & $-0.29^{+0.08}_{-0.05}$
    & $0.729$
    & $0.074$
    \\
$V_2^{BK_{1A}}$
    & $0.41\pm 0.08$
    & $1.51$
    & $1.18~~$
&$V_2^{BK_{1B}}$
    & $-0.17^{+0.05}_{-0.03}$
    & $0.919$
    & $0.855$
    \\
$V_0^{BK_{1A}}$
    & $0.22\pm0.04$
    & $2.40$
    & $1.78~~$
&$V_0^{BK_{1B}}$
    & $-0.45^{+0.12}_{-0.08}$
    & $1.34$
    & $0.690$
    \\
$A^{BK_{1A}}$
    & $0.45\pm0.09$
    & $1.60$
    & $0.974$
&$A^{BK_{1B}}$
    & $-0.37^{+0.10}_{-0.06}$
    & $1.72$
    & $0.912$
    \\
$T_1^{BK_{1A}}$
    & $0.31^{+0.09}_{-0.05}$
    & $2.01$
    & $1.50$
&$T_1^{BK_{1B}}$
    & $-0.25^{+0.06}_{-0.07}$
    & $1.59$
    & $0.790$
    \\
$T_2^{BK_{1A}}$
    & $0.31^{+0.09}_{-0.05}$
    & $0.629$
    & $0.387$
&$T_2^{BK_{1B}}$
    & $-0.25^{+0.06}_{-0.07}$
    & $0.378$
    & $-0.755$
    \\
$T_3^{BK_{1A}}$
    & $0.28^{+0.08}_{-0.05}$
    & $1.36$
    & $0.720$
&$T_3^{BK_{1B}}$
    & $-0.11\pm 0.02$
    & $-1.61$
    & $10.2$
\end{tabular}
\end{ruledtabular}
\end{table}
%
\section{$\barB \to \barkone \ell^+ \ell^-$ decays in the SM}\label{sec:decay}
%
The decay amplitude for $B\to\kone\lpm$ which is analogous to the
$B\to\Kstar\lpm$ decay \cite{Ali:1999mm} is given by
\begin{eqnarray}
{\cal M} &=& \frac{G_F \alpha_{\rm em}}{2\sqrt{2}\pi} V_{ts}^* V_{tb}^{}\,
m_B
\cdot (-i)\left[
  \T_\mu^{(\kone),1} \lbar \gamma^\mu \l
 +\T_\mu^{(\kone),2} \lbar \gamma^\mu \gamma_5 \l
\right],
\end{eqnarray}
where
\begin{eqnarray}
\T_\mu^{(\kone),1} &=&
 \A^\kone(\hats) \epsilon_{\mu\nu\rho\sigma}
 \varepsilon^{*\nu} \hatp_B^\rho \hatp_\kone^\sigma
-i \B^\kone(\hats)\varepsilon^{*}_\mu
\nonumber\\&&
+i \C^\kone(\hats)( \varepsilon^{*} \cdot \hatp_B) \hatp_\mu
+i \D^\kone(\hats)( \varepsilon^{*} \cdot \hatp_B) \hatq_\mu,
\\
\T_\mu^{(\kone),2} &=&
 \E^\kone(\hats) \epsilon_{\mu\nu\rho\sigma}
 \varepsilon^{*\nu} \hatp_B^\rho \hatp_\kone^\sigma
-i \F^\kone(\hats)\varepsilon^{*}_\mu
\nonumber\\&&
+i \G^\kone(\hats)( \varepsilon^{*} \cdot \hatp_B) \hatp_\mu
+i \H^\kone(\hats)( \varepsilon^{*} \cdot \hatp_B) \hatq_\mu,
\end{eqnarray}
with
 $\hatp = p/m_B$,
 $\hatp_B = p_B/m_B$,
 $\hatq= q/m_B$ and
 $p = p_B + p_\kone$,
 $q = p_B - p_\kone = p_+ + p_- $.
Here $\A^\kone(\hats), \cdots, \H^\kone(\hats)$ are defined by
\begin{eqnarray}
\A^\kone(\hats) &=& \frac{2}{1+\hatm_{K_1}} c_9^{\eff} (\hats) A^\kone(\hats) +
\frac{4\hatm_b}{\hats} c_7^\eff T^\kone_1(\hats), \label{Eq:A}
\\
\B^\kone(\hats) &=& (1+\hatm_{K_1})\left[
 c_9^\eff (\hats) V_1^\kone(\hats)
 + \frac{2\hatm_b}{\hats} (1-\hatm_{K_1})c_7^\eff T^\kone_2(\hats)
\right],
\\
\C^\kone(\hats)
&=& \frac{1}{1-\hatm_\kone^2}
\left[
 (1-\hatm_{K_1}) c_9^\eff(\hats) V_2^\kone (\hats) + 2\hatm_b c_7^\eff
 \left(
  T_3^\kone(\hats) + \frac{1-\hatm_\kone^2}{\hats} T_2^\kone(\hats)
 \right)
\right],
\nonumber\\
\\
\D^\kone(\hats)
&=& \frac{1}{\hats}
\biggl[
 c_9^\eff(\hats) \left\{(1+\hatm_\kone) V_1^\kone(\hats)
  - (1-\hatm_\kone) V_2^\kone(\hats)
  - 2\hatm_\kone V_0^\kone(\hats) \right\}
\nonumber\\&&
  - 2\hatm_b c_7^\eff T_3^\kone(\hats)
\biggr],
\\
\E^\kone(\hats) &=& \frac{2}{1+\hatm_\kone} c_{10} A^\kone(\hats),
\\
\F^\kone(\hats) &=& (1 + \hatm_\kone) c_{10} V_1^\kone(\hats),
\\
\G^\kone(\hats) &=& \frac{1}{1 + \hatm_\kone} c_{10} V_2^\kone(\hats),
\\
\H^\kone(\hats)
&=& \frac{1}{\hats} c_{10}
\left[
 (1+\hatm_\kone) V_1^\kone(\hats)
 - (1-\hatm_{K_1}) V_2^\kone(\hats) - 2\hatm_\kone V_0^\kone(\hats)
\right],
\label{Eq:H}
\end{eqnarray}
with $\hatm_\kone = m_\kone/m_B$.  We choose $\hats = \hatq^2$ and $\hatu
\equiv (\hatp_B - \hatp_-)^2 - (\hatp_B - \hatp_+)^2$ as the two independent
parameters, which are bounded as $4\hatm_l^2 \le \hats \le (1-\hatm_{\kone})^2$
and $-\hatu(\hats) \le \hatu \le \hatu(\hats)$, with $\hatu(\hats) \equiv
\sqrt{\lambda \left(1-{4\hatm_l^2}/{\hats}\right)}$, $\lambda \equiv 1 +
\hatm_{\kone}^2 + \hats^2 - 2\hats - 2\hatm_{\kone}^2 (1+\hats)$.  We have
$\hatu = - \hatu(\hats) \cos\theta$, where $\theta$ is the angle between the
momenta of $\l^+$ and the $b$ quark in the center-of-mass frame of the lepton
pair.  We will use the parameters given in Tables~\ref{tab:FFinLF} and
\ref{input} in the numerical analysis.
%
\begin{table}[tbp]
\caption{Input parameters}\label{input}
\begin{center}
\begin{ruledtabular}
\begin{tabular}{l}
$B$ meson mass and lifetimes \cite{Yao:2006px}\\
$m_B=5.279 \GeV$,\quad
$\tau_{B^-}=1.638\times 10^{-12}\,{\rm sec}$,\quad
$\tau_{B^0}=1.530\times 10^{-12}\,{\rm sec}$
\\
\hline
Axial vector meson masses [\GeV]\\
$m_{\konel} = 1.272$  \cite{Yao:2006px},\quad
$m_{\koneh} = 1.403$  \cite{Yao:2006px},\quad
$m_{\konea} = 1.31$ \cite{Yang:2007zt},\quad
$m_{\koneb} = 1.34$ \cite{Yang:2007zt}
\\
\hline
CKM matrix elements \\
$|V_{tb}^{}V_{ts}^*| = 0.0407_{-0.0008}^{+0.0009}$ \cite{CKMfitter}
\\
\hline
$b$ quark mass [\GeV]\\
$m_{b,\rm pole} = 4.8\pm0.2$
\\
\hline
Gauge couplings and the parameter for the $B$ meson distribution
 amplitude \\
  $\alphaem = 1/129$, \quad $\alphas(\mu_h) = 0.3$,
\quad
 $\lambda_{B,+}^{-1} = 3\pm1 \GeV^{-1}$ \cite{Beneke:2001at}
\\
\hline
  $\kone$ decay constants [\MeV] \cite{Yang:2007zt} \\
$f_{\konea}^{\parallel} = 250\pm13$,
\quad
$f_{\koneb}^{\perp}[1\GeV] = 190\pm10$
\\
\hline
Gegenbauer moments at the scale $2.2\GeV$ \cite{Yang:2007zt} \\
$a_0^{\konea,\perp} = \errpm{0.24}{0.03}{0.21}$,\quad
$a_1^{\konea,\perp} = -0.84\pm0.37$,\quad
$a_2^{\konea,\perp} = 0.01\pm0.15$,\quad
\\
$a_1^{\koneb,\perp} = \errpm{0.25}{0.00}{0.26}$,\quad
$a_2^{\koneb,\perp} = -0.02\pm0.17$
\end{tabular}\end{ruledtabular}\end{center}
\end{table}
%
\subsection{Dilepton mass spectrum}\label{sec:spectram-sm}
%
The dilepton invariant mass spectrum of the lepton pair for the
$\barB\to\barkone\lpm$ decay is given by
\begin{eqnarray}
\lefteqn{\frac{d \Gamma(\barB\to\barkone\lpm)}{d \hats}}
\nonumber\\
&=&
\frac{G_F^2 \alphaem^2 m_B^5}{2^{10}\pi^5}
 \left|V_{tb}V_{ts}^*\right|^2 \hatu(\hats)
\times
\Biggl\{
\frac{\left|\A^\kone\right|^2}{3} \hats \lambda
\left(1 + 2\frac{\hatm_\l^2}{\hats}\right)
+ \left|\E^\kone\right|^2 \hats \frac{\hatu(\hats)^2}{3}
\nonumber\\&&
 + \frac{1}{4\hatm_\kone^2}
  \Biggl[
     \left|\B^\kone\right|^2 \left(\lambda - \frac{\hatu(\hats)^2}{3}
   + 8\hatm_\kone^2 (\hats + 2\hatm_\l^2 )\right)
   \nonumber\\&& \phantom{MMMM}
   + \left|\F^\kone\right|^2 \left(\lambda - \frac{\hatu(\hats)^2}{3}
   + 8\hatm_\kone^2 (\hats -4 \hatm_\l^2 )\right)
 \Biggr]
\nonumber\\&&
+ \frac{\lambda}{4\hatm_\kone^2}
\left[
   \left|\C^\kone\right|^2 \left(\lambda - \frac{\hatu(\hats)^2}{3}  \right)
  +\left|\G^\kone\right|^2 \left(\lambda - \frac{\hatu(\hats)^2}{3}
 + 4\hatm_\l^2 (2 + 2\hatm_\kone^2 - \hats)
  \right)
\right]
\nonumber\\&&
 - \frac{1}{2\hatm_\kone^2}
 \Biggl[ \Re \left(\B^\kone \C^{\kone*}\right)
  \left(\lambda - \frac{\hatu(\hats)^2}{3}  \right)
  (1 - \hatm_\kone^2 - \hats)
\nonumber\\&&
 \phantom{MMMM}
 +\Re \left(\F^\kone \G^{\kone*}\right)
  \left(\left(\lambda - \frac{\hatu(\hats)^2}{3}  \right)
  (1 - \hatm_\kone^2 - \hats)+ 4\hatm_\l^2 \lambda\right)
 \Biggr]
\nonumber\\&&
 - 2\frac{\hatm_\l^2}{\hatm_\kone^2} \lambda
 \left[
  \Re \left(\F^\kone \H^{\kone*}\right)
  - \Re \left(\G^\kone\H^{\kone*}\right) (1-\hatm_\kone^2)
 \right]
 + \frac{\hatm_\l^2}{\hatm_\kone^2} \hats \lambda \left|\H^\kone\right|^2
 \Biggr\}.
\end{eqnarray}
%
\begin{figure}[tbp]
\caption{ The dilepton invariant mass distributions for differential decay
rates $d\Br(\Bm\to\konem\mu^+\mu^-)/ds$ in the SM. The central values of inputs
are used.
The solid, dotted and dashed curves correspond to $\thetaK =
-34\degree$, $-45\degree$, $-57\degree$, respectively.
The thick (blue) [thin (red)] curves correspond to values with [without]
resonant corrections.}\label{BRplot}
\begin{center}
\includegraphics[width=3.2in]{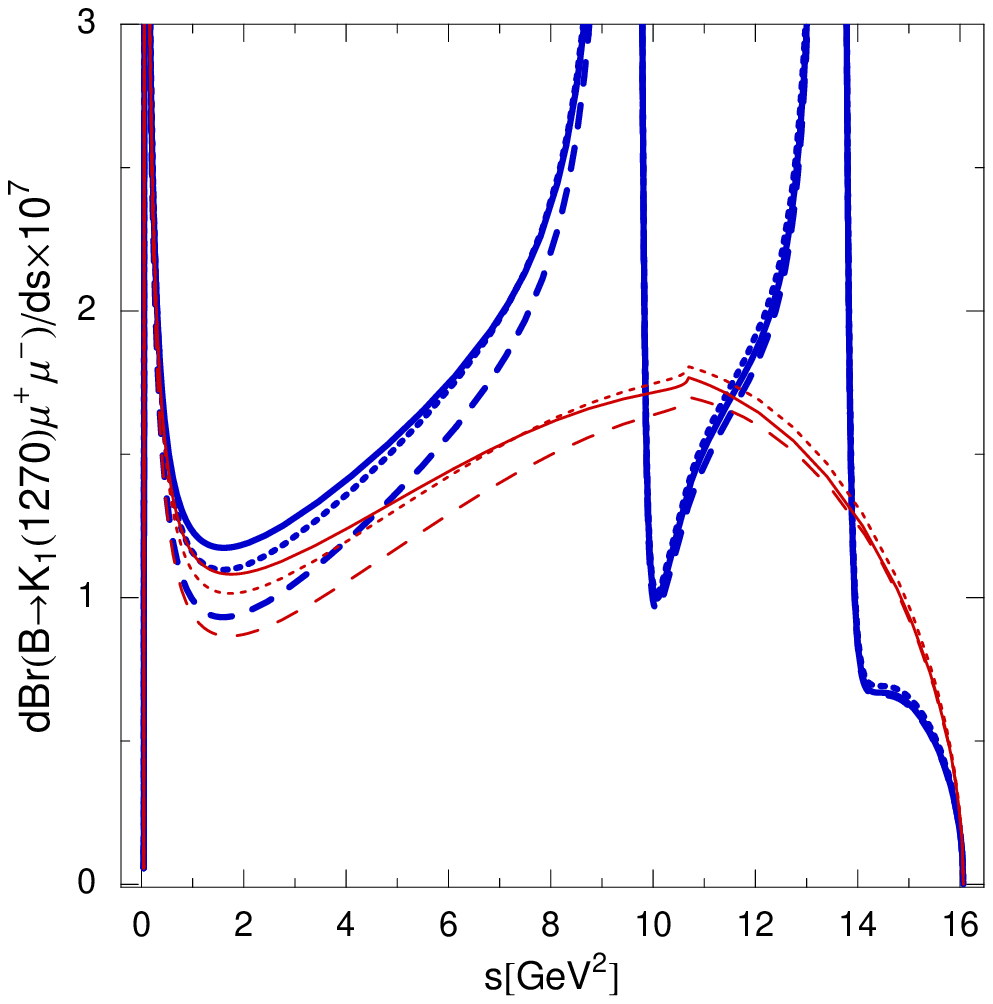}\hfill
\includegraphics[width=3.2in]{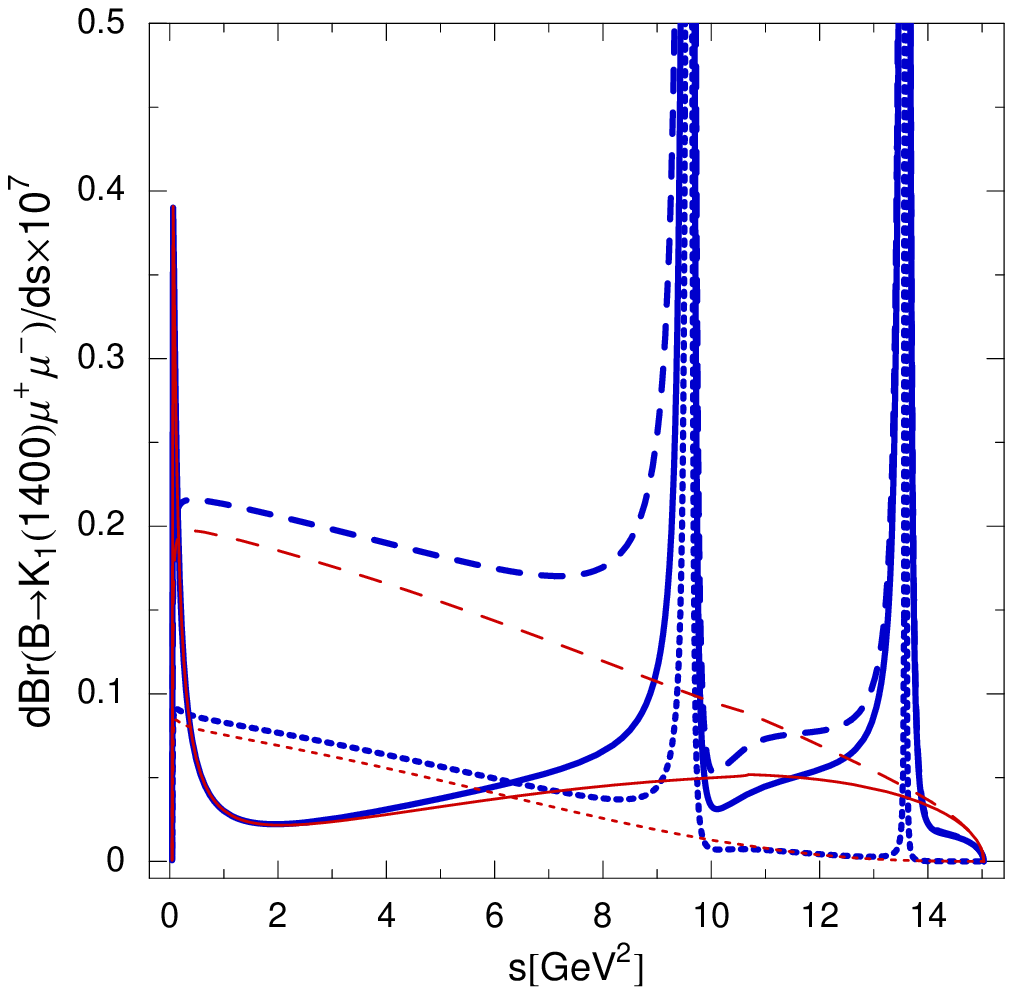}
\end{center}
\end{figure}
%
The differential decay rates $d\Br(\Bm\to \konem \mupm)/ds \equiv \tau_{\Bm}
\times d\Gamma(B\to\konem\mupm)/ds$ are plotted in Fig.~\ref{BRplot}. To
illustrate the dependence on $\thetaK$, we plot the distributions for the
differential decay rates with $\thetaK = -34\degree$, $-45\degree$ and
$-57\degree$, respectively. The effects of charmonium resonances become large
for the large region with $s \gtrsim 5\GeV^2$. We find that in the low $s$
region, where $s\approx 2\GeV^2$, the differential decay rate for $B\to
K_1(1400) \mu^+ \mu^-$ with $\theta_{K_1}=-57^\circ$ is enhanced by about 80\%
compared with that with $\theta_{K_1}=-34^\circ$, whereas the rates for $B\to
K_1(1270) \mu^+ \mu^-$ is not so sensitive to variation of $\theta_{K_1}$.
One should note that the distribution in the low $s$ region is dominated by the
$1/s$ term arising from $B\to K_1 \gamma$; for instance, for the
$B\to\konel\mupm$ decay, it results in the peak at $s\sim 4m_\ell^2$ (or
exactly at $s=0$) and contributes about $-30\%$ at around $s=2\GeV^2$ for
$-57\degree < \thetaK < -34\degree$.

\begin{figure}[tbp]
\caption{ The ratio of the decay distributions, $\RdGamma$ (see the text), as a
function of the dimuon invariant mass $s$.  The legends are the same as in
Fig.~\ref{BRplot}.  }\label{specratio}
\begin{center}
\includegraphics[width=3.2in]{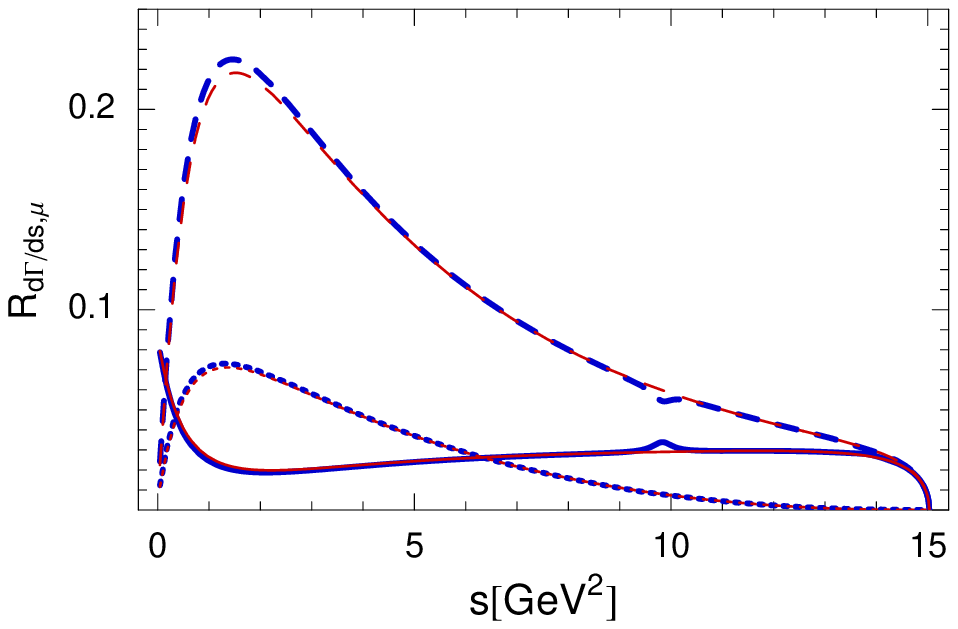}
\end{center}
\end{figure}

Furthermore, the value of $\thetaK$ can be well determined from the following
ratio of the distributions,
\begin{eqnarray}
\RdGamma \equiv
\frac{d\Gamma(\Bm\to\konehm\mupm)/ds}{d\Gamma(\Bm\to\konelm\mupm)/ds}.
\end{eqnarray}
In Fig.~\ref{specratio}, we plot the $\RdGamma$ as a function of $s$, which is
highly insensitive to the resonance contributions and form factors.  When the
magnitude of $\thetaK$ is increased, this ratio peaks at about $s =1.5\GeV^2$
(for $\theta_{K_1}\gtrsim 40^\circ$).

%
\subsection{Branching fractions}
%
\begin{figure}[tbp]
\caption{
Non-resonant branching fractions
$\Br_{\rm nr} (\Bm\to \konem \l^+\l^-)$ as functions of $\thetaK$.
(a) The
thick solid,
thick dashed,
thin solid and
thin dashed
curves correspond to the decays
$B \to \konel \epm$,
$\koneh \epm$,
$\konel \mupm$ and
$\koneh \mupm$, respectively.
(b) The
solid and dashed curves correspond to
$B \to \konel \tau^+ \tau^-$,
$B \to \koneh \tau^+ \tau^-$,
respectively.
The vertical lines indicate the allowed range of $\thetaK$ given in
Eq.~(\ref{thetaKvalue}) \cite{Hatanaka:2008xj}.}\label{BRplotThetaK}
\begin{center}
\includegraphics[width=3.2in]{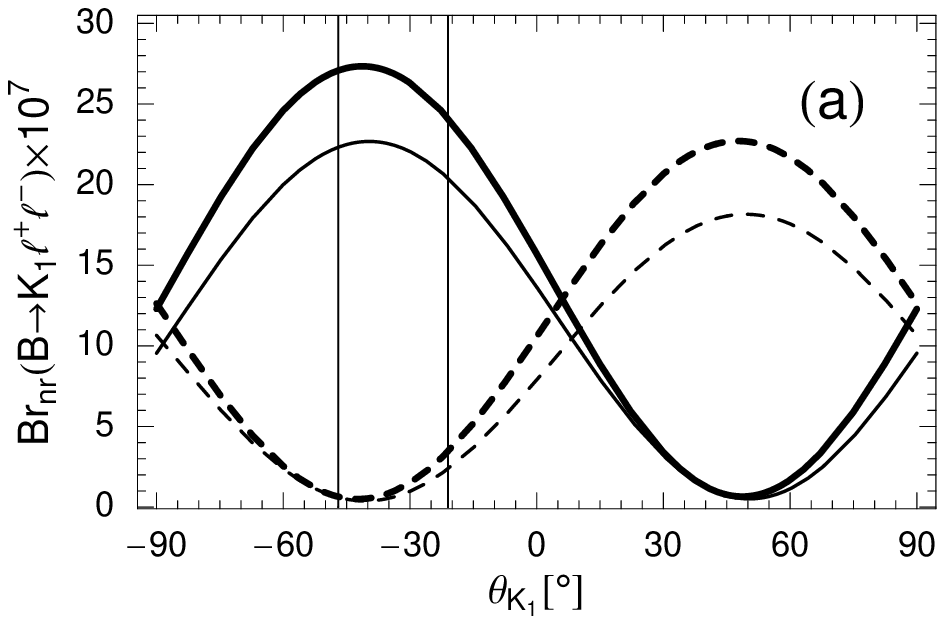}
\includegraphics[width=3.2in]{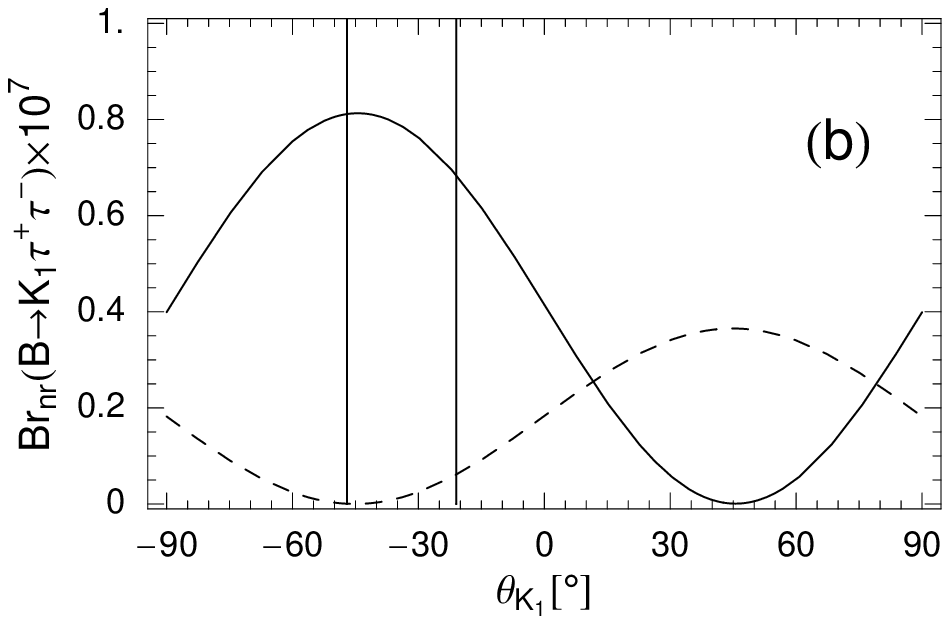}
\end{center}
\end{figure}
%
\par 
%
\begin{table}[tbp]
\caption{ Predictions for the non-resonant branching fractions $\Br_{\rm
nr}(B \to \kone\lpm$).  The first and second errors come from the
uncertainty of the form factors and of the $\thetaK$ within the allowed
region \cite{Hatanaka:2008xj}, respectively. }\label{kll-predict}
\begin{center}
\begin{ruledtabular}
\begin{tabular}{ll|ll}
Mode                   & $\Br_{\rm nr} \times 10^6$ &
Mode                   & $\Br_{\rm nr} \times 10^6$ \\
\hline
$\Bm\to\konelm\epm$    & $\errpmf{2.7}{1.5}{1.2}{0.0}{0.3}$   &
$\Bz\to\konelz\epm$    & $\errpmf{2.5}{1.4}{1.1}{0.0}{0.3}$
\\
$\Bm\to\konelm\mupm$   & $\errpmf{2.3}{1.3}{1.0}{0.0}{0.2}$   &
$\Bz\to\konelz\mupm$   & $\errpmf{2.1}{1.2}{0.9}{0.0}{0.2}$
\\
$\Bm\to\konelm\taupm$  & $\errpmf{0.08}{0.04}{0.03}{0.00}{0.01}$   &
$\Bz\to\konelz\taupm$  & $\errpmf{0.08}{0.04}{0.03}{0.00}{0.01}$
\\
$\Bm\to\konehm\epm$    & $\errpmf{0.10}{0.03}{0.03}{0.25}{0.05}$   &
$\Bz\to\konehz\epm$    & $\errpmf{0.09}{0.03}{0.03}{0.23}{0.04}$
\\
$\Bm\to\konehm\mupm$   & $\errpmf{0.06}{0.02}{0.01}{0.18}{0.02}$   &
$\Bz\to\konehz\mupm$   & $\errpmf{0.06}{0.02}{0.01}{0.18}{0.02}$
\\
$\Bm\to\konehm\taupm$  & $\errpmf{0.001}{0.000}{0.000}{0.005}{0.001}$ &
$\Bz\to\konehz\taupm$  & $\errpmf{0.001}{0.000}{0.000}{0.005}{0.001}$
\end{tabular}
\end{ruledtabular}
\end{center}
\end{table}
%

\begin{figure}[tbp]
\caption{ $R_{\l, \rm nr} \equiv \Br_{\rm nr}(B\to\koneh\l^+\l^-)/\Br_{\rm
nr}(B\to\konel\l^+\l^-)$ as functions of $\thetaK$. The solid, dashed and
dot-dashed curves correspond to $R_{e, \rm nr}$, $R_{\mu, \rm nr}$ and
$R_{\tau, \rm nr}$, respectively. The allowed range of $\thetaK$ given in
Eq.~\eqref{thetaKvalue} \cite{Hatanaka:2008xj} is also shown. }
\label{brratioplot}
\begin{center}
\includegraphics[width=3.2in]{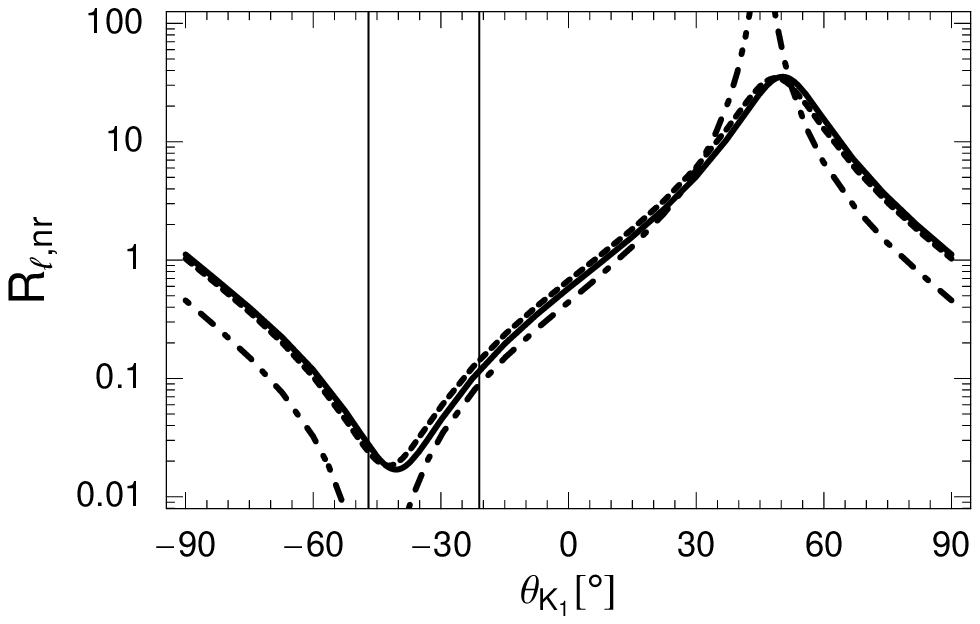}
\end{center}
\end{figure}

In Table~\ref{kll-predict}, we summarize the predictions for branching
fractions corresponding to $\thetaK=-(34\pm13)\degree$. The branching fractions
for $B\to\kone\epm$ and $B\to\kone\mupm$ are close to $B\to\Kstar\epm$,
$B\to\Kstar\mupm$ given in \cite{Ali:1999mm}.  On the other hand, the branching
fractions for $B \to \kone \tau^+ \tau^-$ decays are very small since the
allowed phase space is quite narrow. In Fig.~\ref{BRplotThetaK}, we plot the
non-resonant branching fractions $\Br_{\rm nr}(\Bm\to\konem \lpm)$ as functions
of $\thetaK$.  For the range of $\thetaK = -(34 \pm 13)\degree$, we obtain
$\Br_{\rm nr}(B \to \konel \lpm) \gg \Br_{\rm nr}(B\to\koneh\lpm)$.  It should
be helpful to define the ratio,
\begin{eqnarray}
R_{\l, \rm nr} \equiv \frac{
\Br_{\rm nr}(B\to \koneh \lpm)}{\Br_{\rm nr} (B\to \konel \lpm)}.
\end{eqnarray}
We show $R_{\l,\rm nr}$ as functions of the $\thetaK$ in
Fig.~\ref{brratioplot}.  These ratios sensitively depend on $\thetaK$, and are
smaller than $0.15$ for $-47\degree \le \thetaK \le -21\degree$.  We predict
\begin{eqnarray}
R_{e, \rm nr} = \errpmf{0.04}{0.01}{0.01}{0.11}{0.02}, \quad \Rmunr =
\errpmf{0.03}{0.01}{0.01}{0.09}{0.01}, \quad R_{\tau,\rm nr} =
\errpmf{0.02}{0.01}{0.00}{0.07}{0.02},
\end{eqnarray}
where the first and second errors correspond to the uncertainties of the form
factors and $\theta_{K_1}$, respectively.   In Fig.~\ref{BrRatioNP}, we will
further show that the ratio $\Rmunr$ is highly insensitive to the NP
corrections.

%

%
\subsection{Forward-backward asymmetry}

The differential forward-backward asymmetry of the
$\barB\to\barkone\lpm$ decay is defined by
\begin{eqnarray}
\frac{d \AFB}{d \hats} \equiv
 \int_0^{ \hatu(\hats)} \! d\hatu \frac{d^2 \Gamma}{d\hatu d\hats}
- \int_{-\hatu(\hats)}^0 \! d\hatu \frac{d^2\Gamma}{d\hatu d\hats},
\end{eqnarray}
which can be written in terms of quantities in
 Eqs.~\eqref{Eq:A}-\eqref{Eq:H} as
\begin{eqnarray}
\frac{d \AFB}{d\hats}
&=& -\frac{G_F^2 \alphaem^2 m_B^5}{2^{10}\pi^5}
|V_{ts}V_{tb}^*| \hats \hatu(\hats)^2
\left\{ \Re\left(\B^\kone \E^{\kone*}\right)
 + \Re\left(\A^\kone \F^{\kone*}\right) \right\},
\end{eqnarray}
and, after including the hard spectator correction \cite{Beneke:2001at}, are
given by
\begin{eqnarray}
\frac{d \AFB}{d\hats}
&=& -\frac{G_F^2 \alphaem^2 m_B^5}{2^8 \pi^5} |V_{ts}^* V_{tb}|^2
 \hats \hatu(\hats)^2 \times
 c_{10}
\Bigg[
 \Re(c_9^\eff(\hats)) A^\kone V_1^\kone
\nonumber\\&& \phantom{MM}
 + \frac{\hatm_b}{\hats} c_7^\eff
\bigg\{
 A^\kone T_2^\kone(1-\hatm_\kone) + V_1^\kone T_1^\kone (1+ \hatm_\kone)
\bigg\} + \frac{\hatm_b}{\hats} \Delta_{\rm HS}
\Bigg],
\end{eqnarray}
where $\Delta_{\rm HS}$ is the hard spectator correction given by
\begin{eqnarray}
\Delta_{\rm HS}
&=&
\bigg\{(1+\hatm_{\kone}) V_1^{\kone}
 + (1-\hatm_{\kone})(1-\hats)A^{\kone}\bigg\}
\nonumber\\&& \phantom{M}\times
\frac{\alpha_s(\mu_h) C_F}{4\pi}\frac{\pi^2}{N_c}
\frac{f_B f_{\kone}^{\perp}}{\lambda_{B,+} m_B} \int_0^1 \!\!\! du\,
\Phi_{\kone}^{\perp}(u) T^{(\rm nf)}_{\perp,+}(u),. \label{AFB}
\end{eqnarray}
Here $f_{\kone}^\perp$ and $\Phi_{\kone}^{\perp}(u)$ are the transverse decay
constant and the twist-2 tensor light-cone distribution amplitude of the
$\kone$, respectively.
$f_{\konel}^\perp$, $f_{\koneh}^\perp$ and $\Phi_{\konel}^\perp$,
$\Phi_{\koneh}^\perp$ are related with $f_{\konea}^\perp$, $f_{\koneb}^\perp$
and $\Phi_{\konea}^\perp$, $\Phi_{\koneb}^\perp$ by \cite{Yang:2007zt}
\begin{eqnarray}
\pmatrix{
f_{\konel}^\perp  \cr
f_{\koneh}^\perp  }
&=& M \cdot \pmatrix{
f_{\konea}^\perp a_0^{\konea,\perp} \cr
f_{\koneb}^\perp    },
\\
\pmatrix{
f_{\konel}^\perp \Phi_{\konel}^\perp \cr
f_{\koneh}^\perp \Phi_{\koneh}^\perp }
 &=& M \cdot
\pmatrix{ f_{\konea}^\perp \Phi_{\konea}^\perp \cr f_{\koneb}^\perp
\Phi_{\koneb}^\perp },
\end{eqnarray}
where  $\Phi_{\konea}^\perp$ and $\Phi_{\koneb}^\perp$ are expanded as
\newcommand{\ubar}{\bar{u}}
\begin{eqnarray}
\Phi_{\konea}^\perp(u) &=&
6u\ubar
\left[a_0^{\konea,\perp} + 3 a_1^{\koneb,\perp} \xi + a_2^{\koneb,\perp}
 \frac{3}{2}(5\xi^2-1)
\right],
\\
\Phi_{\koneb}^\perp(u) &=&
6u\ubar
\left[1 + 3 a_1^{\koneb,\perp} \xi + a_2^{\koneb,\perp}
 \frac{3}{2}(5\xi^2-1)
\right],
\end{eqnarray}
with $a_0^{K_{1B},\perp} \equiv 1$, $\ubar \equiv 1-u$ and $\xi \equiv u -
\ubar$.  The values of $f_{\konea}^\perp$, $f_{\koneb}^\perp$ and the
Gegenbauer moments, $a_i^{\kone,\perp}$, are tabulated in Table~\ref{input}.

\begin{figure}[tbp]
\caption{%
Normalized differential forward-backward asymmetries: (a) $d\barAFB(\Bm \to
\konelm \mupm)/ds$ and (b) $d\barAFB(\Bm \to \konehm \mupm)/ds$. The legends
are the same as Fig.~\ref{BRplot}.  }\label{AFBplot}
\begin{center}
\includegraphics[width=3.2in]{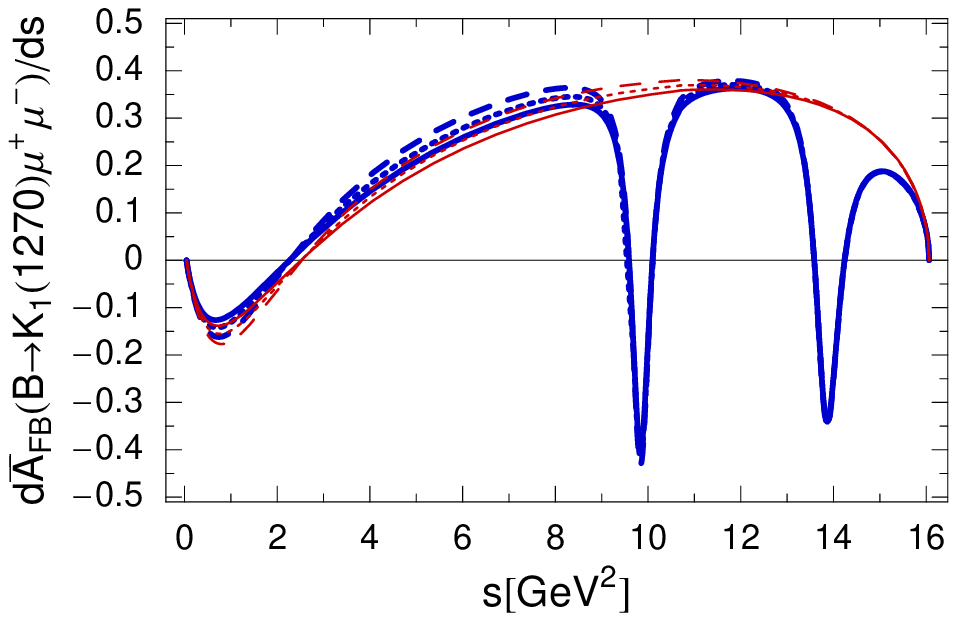}\hfill
\includegraphics[width=3.2in]{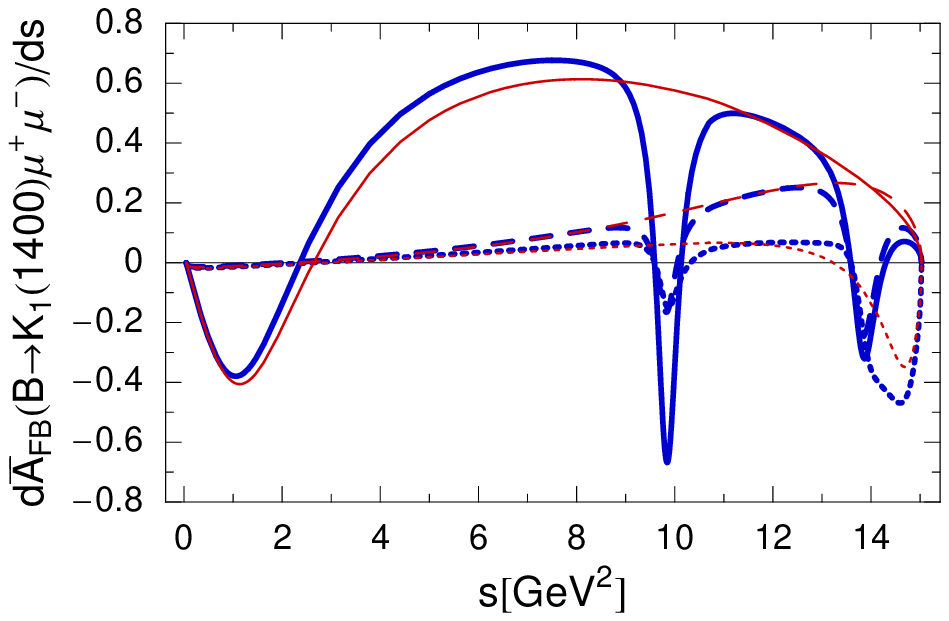}%
\end{center}
\end{figure}

In the following, to compare the theoretical predictions with the data, we use
the normalized differential forward-backward asymmetry as
\begin{eqnarray}
\dfrac{d\barAFB}{d \hats}
\equiv
\frac{d \AFB}{d \hats}
\Bigg/
\frac{d\Gamma}{d \hats}.
\end{eqnarray}
In Fig.~\ref{AFBplot}, the normalized differential forward-backward asymmetries
$d\barAFB(\Bm\to\konem\mupm)/ds$ versus $s$ are plotted. For $B\to \konel\mupm$
decays, the dependence of $d\barAFB/ds$ on $\thetaK$ is negligibly small. For
$\thetaK \lesssim -45\degree$, $d\barAFB(B\to\koneh\mupm)/ds$ almost vanishes
in the region below the $J/\psi$ resonance.  We define $s_0^\kone$ to be the
position of zero of the FBA. $s_0^{K_1}$ satisfies
\begin{eqnarray}\label{eq:c0}
 \frac{\Re(c_9^\eff(\hats_0^\kone)) }{ c_7^{\eff,{\rm HS}}(\hats_0^{\kone})}
&=&
 - \frac{\hatm_b}{\hats_0^\kone}
\left\{
 \frac{T_2^\kone(\hats_0^\kone)}{V_1^\kone(\hats_0^\kone)}
 (1-\hatm_\kone)
+\frac{T_1^\kone(\hats_0^\kone)}{A^\kone  (\hats_0^\kone)}(1+\hatm_\kone)
\right\},
\end{eqnarray}
which is negative.
Here
\begin{eqnarray}
c_7^{\eff,{\rm HS}}(\hats)
&\equiv& c_7^{\eff}
 + \frac{\Delta_{\rm HS}(\hats)}{
A^{\kone}(\hats) T_2^{\kone}(\hats) (1-\hatm_{\kone})
 + V_1^{\kone}(\hats) T_1^{\kone}(\hats)(1+\hatm_\kone)}.
\end{eqnarray}
The position of zero appears below the $J/\psi$-resonance region and depends
weakly on $\thetaK$, especially for $B \to K_1(1270) \lpm$ as shown in
Fig.~\ref{AFBplot}. We obtain the positions of the zeros of forward-backward
asymmetries to be
\begin{eqnarray}
s_0^{\konel} = \errpmf{2.27}{0.04}{0.07}{0.01}{0.01}\GeV^2
\quad\mbox{ and }\quad
s_0^{\koneh} =\errpmf{2.80}{0.23}{0.29}{0.74}{0.07}\GeV^2,
\end{eqnarray}
where the first and second errors correspond to the uncertainties of the form
factors and $\thetaK (=-(34\pm 13)^\circ)$, respectively. In the following
section, we will show that, as the $B \to \Kstar \lpm$ decay, for the
$B\to\konel\lpm$ decay the position of the zero of the FBA can be a good
observable for searching for new-physics effects.
%

%

\section{NP effects}\label{sec:NP}
%
In this section, we study the NP corrections to the $B^- \to \konem \mupm$
decays in the model-independent way. As in Ref.~\cite{Ali:1999mm}, we
parametrize the NP contributions to the Wilson coefficients as
\begin{eqnarray}
c_i \equiv c_i^{\SM} + c_i^{\NP} = R_i \, c_i^{\SM}
\quad\mbox{ for }\quad c_i = c_7^{\eff}, c_9, c_{10},
\end{eqnarray}
at the scale $\overline{m}_b(\overline{m}_b)$.  For simplicity, we assume all
$R_{i}$ are real. The model-independent analysis for $B\to X_s \gamma$ and
$B\to X_s \lpm$ \cite{Ali:2002jg} gives the following constraints,
\begin{eqnarray}
0.8 \lesssim |R_7| \lesssim 1.2,
\quad
1 \lesssim R_9^2 + R_{10}^2 \lesssim 4.
\end{eqnarray}
The possibility of flipped sign of $c_7^{\rm eff}$ due to the NP contribution
in the minimum supersymmetric standard model (MSSM) with the minimal flavor
violation (MFV) ansatz and with large $\tan\beta$ has been studied in
Ref.~\cite{Feldmann:2002iw}. The twofold constraint was given by
\begin{eqnarray}
-0.02 \le c_7^{\NP}\le 0.12 \quad\mbox{or}\quad
 0.59 \le c_7^{\NP} \le 1.24,
\end{eqnarray}
at the weak scale. Further constraints on $c_7^{\NP}$ have been obtained with
\begin{eqnarray}
c_7^{\NP}
 &=&  -0.039 \pm 0.043 \cup 0.931 \pm 0.016 \quad\mbox{($68\%$ CL)} \\
 &=&  [-0.104,0.026] \cup [0.874,0.988]    \quad\mbox{($95\%$ CL)}
\end{eqnarray}
  in Ref.~\cite{Haisch:2007ia} and
\begin{eqnarray}
c_7^{\NP}
 &=& 0.02 \pm 0.047 \cup 0.958 \pm 0.002 \quad\mbox{($68\%$ CL)} \\
 &=& [-0.039,0.08] \cup [0.859,1.031]\quad \quad\mbox{($95\%$ CL)}
\end{eqnarray}
in Ref.~\cite{Bobeth:2005ck}. The sign of Re($c_7^{\rm eff}$) can also
be flipped in supersymmetric models with non-minimal flavor violation
via gluino-down-squark loops. Furthermore, in general flavor-violating
supersymmetric models the sign of $c_9$ and $c_{10}$ can be
flipped. Therefore, in the present paper, we consider $R_i = 1.2$
(i.e. $20\%$ enhancement for the SM Wilson coefficients due to the NP
correction), $1.0$ (i.e. without the NP correction), $0.8$ (i.e. $20\%$
smaller than the SM Wilson coefficients) and $-1.0$ (i.e. the Wilson
coefficients are in opposite signs but have the same magnitudes compared
to the SM results).

\begin{figure}[tbp]
\caption{ $\Rmunr=\Br_{\rm nr}(B\to\koneh\mupm) / \Br_{\rm
nr}(B\to\konel\mupm)$ as a function of $\thetaK$. Variations of NP with
$(R_7,R_9,R_{10}) = (r,1,1),(1,r,1),$ and $(1,1,r)$ are respectively included,
where $r = 1.0$ (solid), $1.2$ (dotted), $0.8$ (dot-dashed) and $-1.0$
(dashed). The vertical lines indicate the allowed range of $\thetaK$ given in
Eq.~\eqref{thetaKvalue} \cite{Hatanaka:2008xj}.}\label{BrRatioNP}
\begin{center}
\includegraphics[width=3.2in]{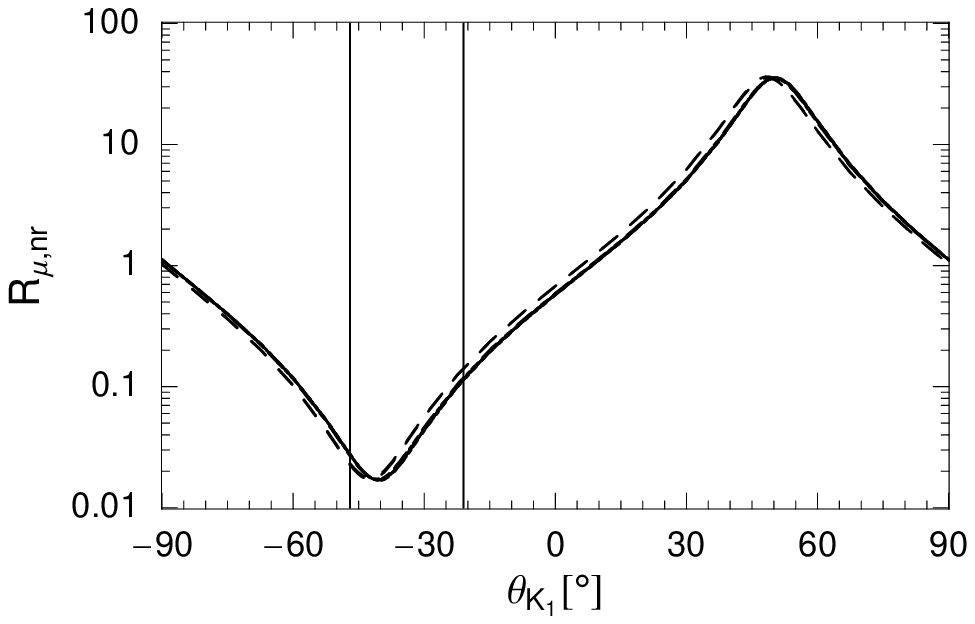}
\end{center}
\end{figure}

\begin{figure}[tbp]
\caption{ $\RdGamma$, the ratio of the differential decay rates, as a function
of the dimuon invariant mass, $s$.
Variations of $R_7$, $R_9$ and $R_{10}$ are depicted in (a), (b) and (c)
respectively, where the remaining $R_i$ are set to their SM values.
The thick (blue) and thin (red) curves correspond to
$\thetaK = -34\degree$ and $-57\degree$, respectively.
The solid, dotted, dot-dashed and dashed curves   correspond to $R_i =
1.0$, $1.2$, $0.8$ and $-1.0$, respectively.  }\label{specRatioNP}
\includegraphics[width=3.2in]{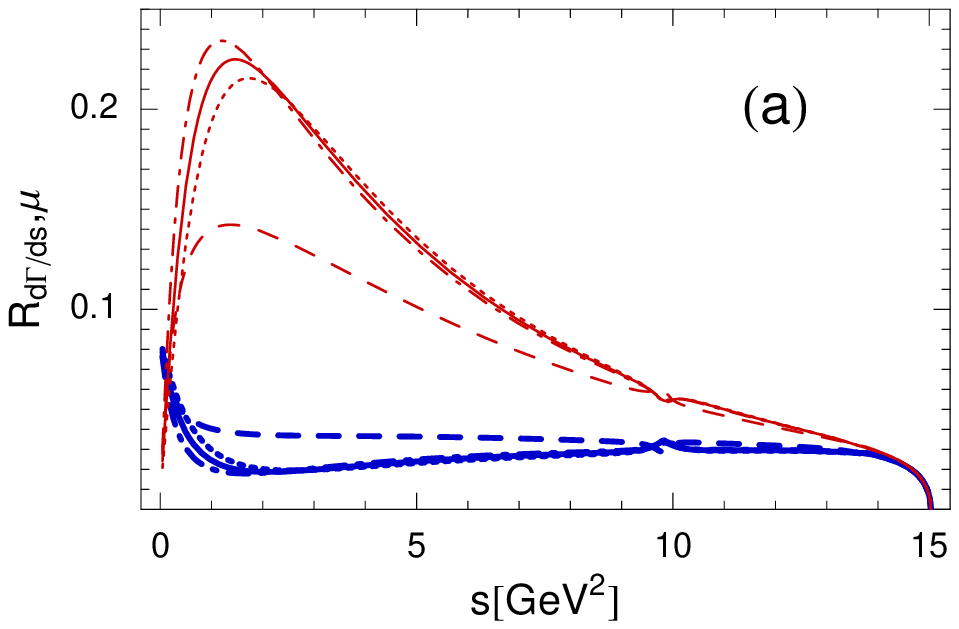}
\includegraphics[width=3.2in]{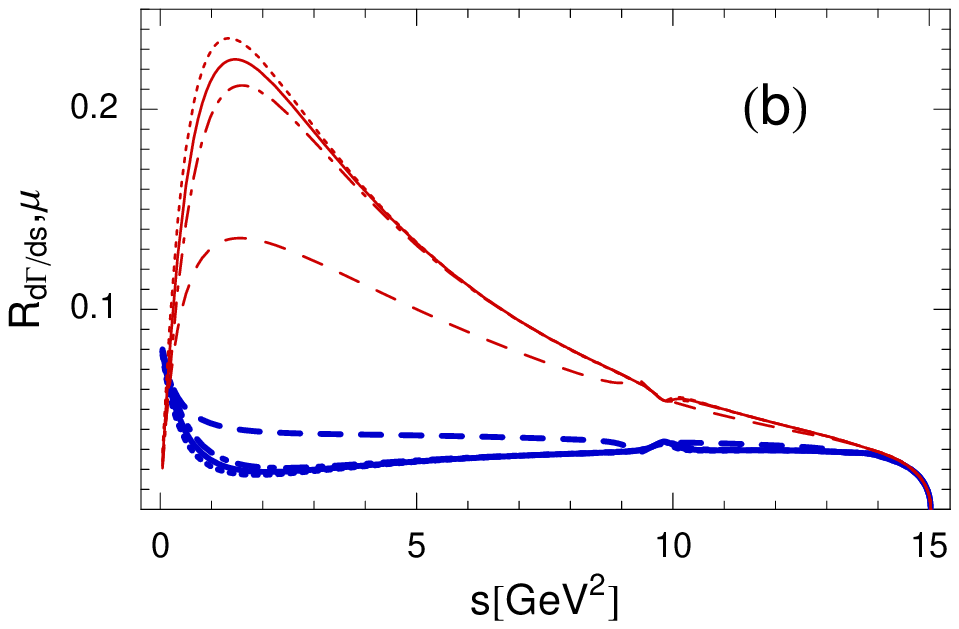}
\includegraphics[width=3.2in]{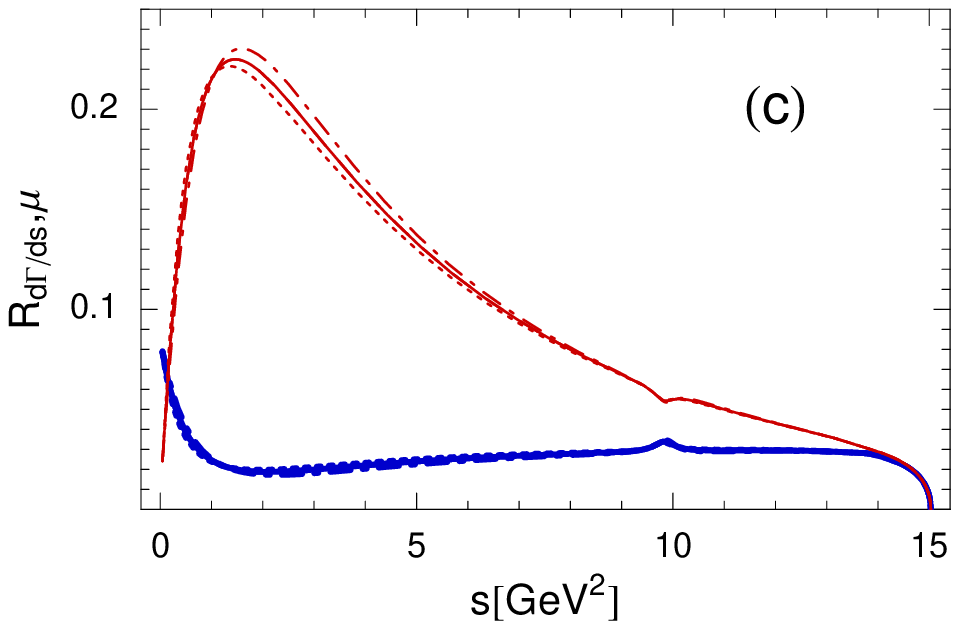}
\end{figure}

In Fig.~\ref{BrRatioNP}, the ratio of the non-resonant branching
fractions $\Rmunr \equiv \Br_{\rm nr}(B\to\koneh\mupm) / \Br_{\rm
nr}(B\to\konel\mupm)$, including the NP corrections, as a function of
the value of $\thetaK$ is depicted. We show that $\Rmunr$ is highly
insensitive to the NP effect and thus is suitable for determining the
value of $\thetaK$. In Fig.~\ref{specRatioNP}, we plot $\RdGamma$, the
ratio of the differential decay rates, as a function of the dimuon
invariant mass, $s$, where the NP effects are considered. We find that
$\RdGamma$ is insensitive to variation of $R_{10}$, whereas its value is
increased (decreased) by about $100\%$ ($40\%$) at about $s=1.5\GeV^2$
corresponding to $\thetaK=-34\degree$ ($-57\degree$) when $R_7$ or $R_9$
equals to $-1$.
%
%

\begin{figure}[tbp]
\caption{%
Normalized differential forward-backward asymmetry
$d\barAFB(B^-\to\konelm\mupm)/ds$ as a function of the dimuon invariant mass
$s$.
The thick (blue) and thin (red) curves correspond to the asymmetries with
$\thetaK=-34\degree$ and $-57\degree$, respectively.
In (a), where $R_9 = R_{10} = 1.0$ (the SM result), the solid curves are for
$R_7 = 1.0$, the dotted for $R_7 = 1.2$, the dot-dashed for $R_7 = 0.8$ and the
dashed for $R_7 = -1.0$.
In (b), where $R_7 = R_{10} = 1.0$ (the SM result), and the solid curves are
for $R_9 = 1.0$, the dotted for $R_9 = 1.2$, the dot-dashed for $R_9=0.8$ and
the dashed for $R_9 = -1.0$.
In (c), where $R_7 = R_9 = 1.0$ and the solid curves are for $R_{10} = 1.0$
(the SM result), the dotted for $R_{10} = 1.2$, the dot-dashed for $R_{10} =
0.8$ and the dashed for $R_{10} = -1.0$.
The $d\barAFB/ds$ with $(R_7,R_9,R_{10})=(1.0,-1.0,-1.0)$,   and
$(-1.0,1.0,-1.0)$ are denoted by the double-dot dashed and long-short dashed
curves, respectively, in (c).}\label{AFB-NP}
\includegraphics[width=3.2in]{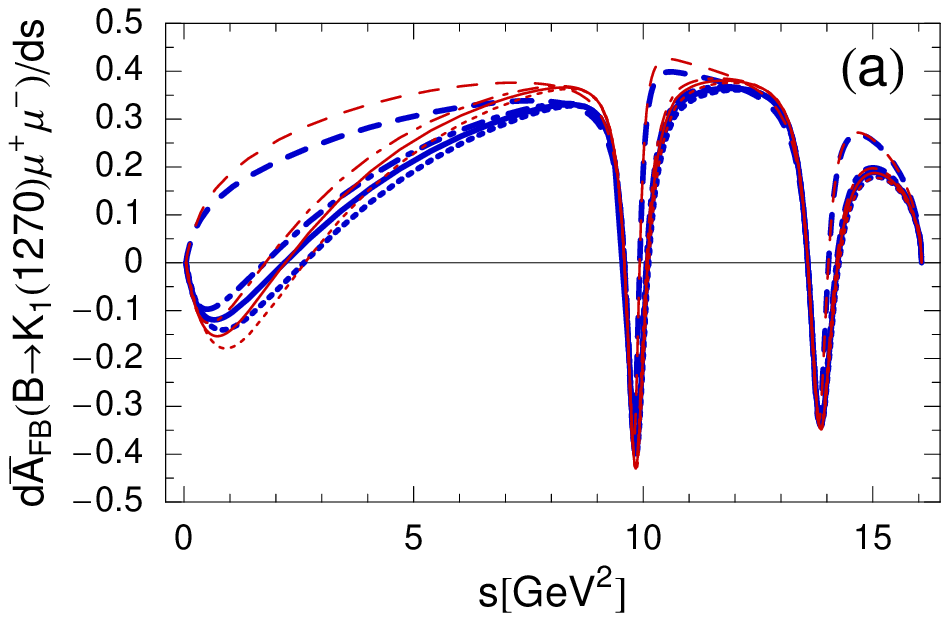}
\includegraphics[width=3.2in]{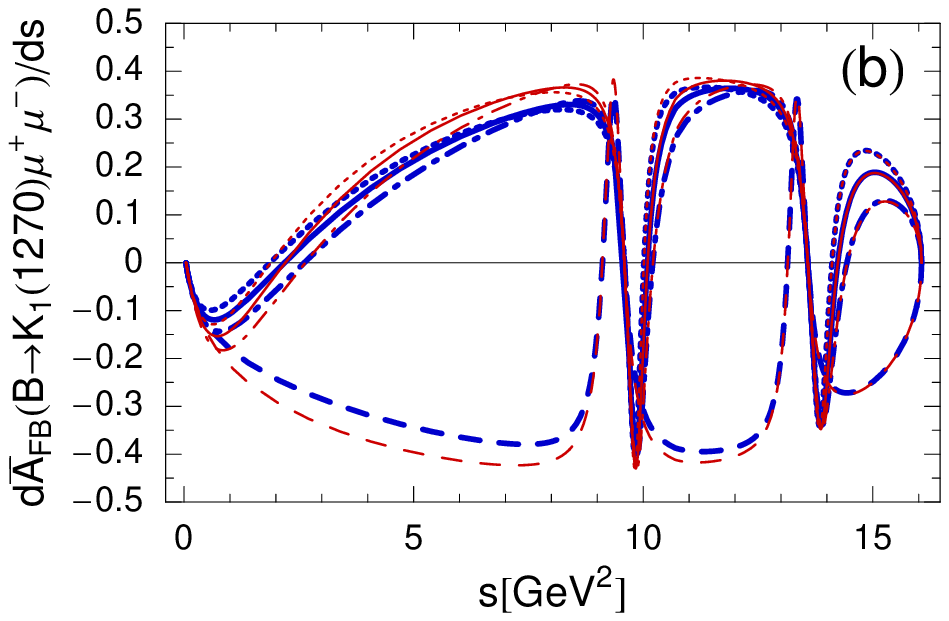}
\includegraphics[width=3.2in]{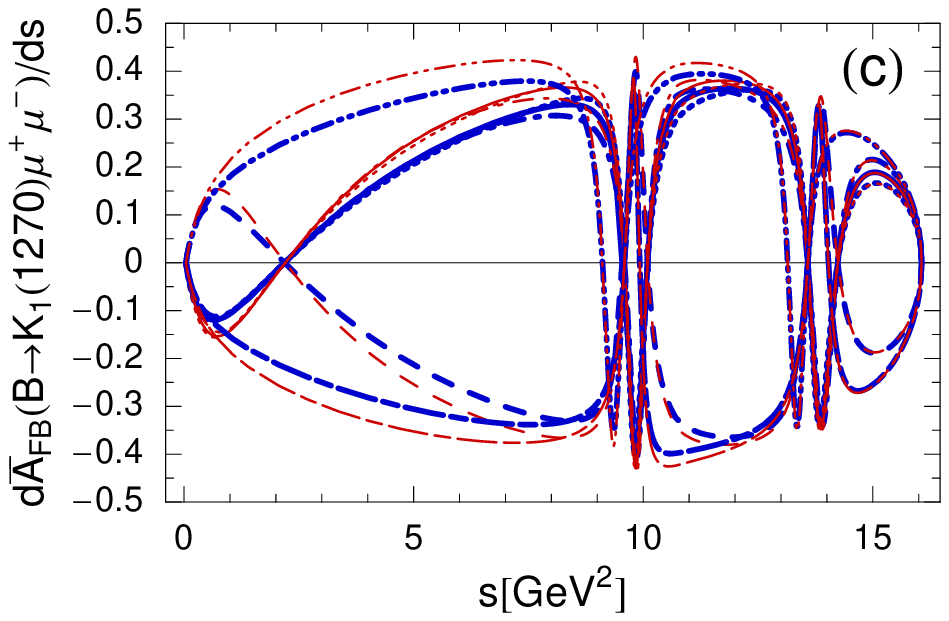}
\end{figure}

Taking into account the possible NP corrections, we plot $d\barAFB(\Bm \to
\konelm\mupm)/ds$ as a function of $s$ in Fig.~\ref{AFB-NP}. We do not consider
the $\Bm\to \konehm \mupm$ decay, since its branching fraction is relatively
small. As shown in Fig.~\ref{AFBplot} (see also Fig.~\ref{AFB-NP}), the
differential forward-backward asymmetry for $B\to\konel\mupm$ and its
$s_0^{\konel}$ (if existing) are very insensitive to variation of $\thetaK$.
For the cases with $c_7^{\rm eff}, c_9$ and $c_{10}$ of SM-like sign, the
change of the FBA zero owing to variation of NP parameters could be manifest as
compared to the hadronic uncertainties. As is well known in the case of $B\to
K^*(892) \ell^+ \ell^-$, for the flipped sign of $c_7^{\rm eff}, c_9$ or
$c_{10}$ the characteristic features of the FBA change dramatically. Because
the asymmetry zero exists only for $\Re(c_9^\eff)/c_7^\eff < 0$ (see
Eq.~(\ref{eq:c0})), therefore there is no asymmetry zero for
$(R_7,R_9)=(\pm1,\mp1)$ in the spectrum. Flipping the sign of $c_{10}$ would
change the sign of the FBA. From the above discussions we can conclude that the
position of the FBA zero for the $B\to\konel\lpm$ decay is a suitable quantity
to constrain the NP parameters. Recent measurements for $B\to\Kstar\lpm$ decays
\cite{Ishikawa:2006fh,aubert:2008ju} seem to favor (i) the flipped sign of
$c_7^{\eff}$ which is denoted by the dashed curves in Fig.~\ref{AFB-NP}(a), or
(ii) the simultaneous flip of the sign of $c_9$ and $c_{10}$ which are denoted
by the double-dot dashed curves in Fig.~\ref{AFB-NP}(c). However, they disfavor
the flipped sign($c_9 c_{10}$) models. See also the discussion in
Ref.~\cite{Bobeth:2008ij}.
%
%
%

\section{Summary}\label{sec:summary}
%
We have studied the rare decays $B \to K_1 \ell^+ \ell^-$ with $K_1
\equiv K_1(1270)$, $K_1(1400)$ and $\ell \equiv e$, $\mu$, $\tau$. The
strange axial-vector mesons, $K_1(1270)$ and $K_1(1400)$, are the
mixtures of the $K_{1A}$ and $K_{1B}$, which are the $1^3P_1$ and
$1^1P_1$ states, respectively.  Although the branching ratios depend on
the magnitudes of $B \to K_1$ form factors, the
$K_1(1270)$--$K_1(1400)$ mixing angle, $\theta_{K_1}$, can be well
determined from the measurement of the ratio $R_\ell \equiv {\cal B}(B
\to K_1(1400)\ell^+\ell^-)/{\cal B}(B \to K_1(1270)\ell^+\ell^-)$, which
depends very weakly on new-physics corrections.  We have calculated
differential forward-backward asymmetries of $B\to K_1\mu^+\mu^-$
decays. For $B \to K_1(1270)\mu^+\mu^-$, the asymmetry zero, which
depends very weakly on $\theta_{K_1}$, can be dramatically changed due
to variation of new-physics parameters.

%
%
\begin{acknowledgments}
This research was supported in part by the National Science Council of
R.O.C. under Grant No.\ NSC96-2112-M-033-004-MY3 and No.\
NSC96-2811-M-033-004.
\end{acknowledgments}
%

\end{document}